# Beyond Global Metrics: A Fairness Analysis for Interpretable Voice Disorder Detection Systems

Mariel Estevez[1], Cyntia Bonomi[1], Dayana Ribas[2,3], Alfonso Ortega[3] and Luciana Ferrer[1]

[1]Instituto de Investigación en Ciencias de la Computación, UBA-CONICET, Argentina

[2]BTS-Business Telecommunications Services

[3]ViVoLab, Aragón Institute for Engineering Research (I3A), University of Zaragoza, Spain

**Purpose:** To investigate and quantify demographic-dependent biases in Automatic Voice Disorders Detection (AVDD) systems by analysing performance disparities across speaker groups, and to evaluate group-specific calibration strategies for improving reliability.

**Method:** We conducted a comprehensive analysis of an AVDD system using existing voice disorder datasets with available demographic metadata. The study involved analysing system performance across various demographic groups, particularly focusing on gender and age-based cohorts. Performance evaluation was based on multiple metrics including normalised costs and cross-entropy. We employed calibration techniques trained separately on predefined demographic groups to address group-dependent miscalibration.

**Results:** Analysis revealed significant performance disparities across demographic groups despite strong global metrics. The system showed systematic biases, misclassifying healthy speakers over 55 as having a voice disorder and speakers with disorders aged 14-30 as healthy. Group-specific calibration improved posterior probability quality, reducing overconfidence. For young disordered speakers, low severity scores were identified as contributing to poor system performance. For older speakers, age-related voice characteristics and potential limitations in the pretrained Hubert model used as feature extractor likely affected results.

**Conclusions:** The study demonstrates that global performance metrics are insufficient for evaluating AVDD system performance. Group-specific analysis may unmask problems in system performance which are hidden within global metrics. Further, group-dependent calibration strategies help mitigate biases, resulting in a more reliable indication of system confidence. These findings emphasize the need for demographic-specific evaluation and calibration in voice disorder detection systems, while providing a methodological framework applicable to broader biomedical classification tasks where demographic metadata is available.

1. **Introduction**

Voice disorders are highly prevalent, with 30% of individuals experiencing at least one period of vocal impairment during their lifetime in a study by Roy et al. (2005). Particularly, these disorders greatly impact those whose voice serves as a primary tool for their professional life. Moreover, individuals facing such issues often defer seeking specialised assistance until the disorder has significantly progressed. Prolonged or untreated voice disorders can lead to diminished quality of life, reduced productivity, and, in severe cases, long-term vocal damage.

In the last few years, there has been a growing interest in the development of Automatic Voice Disorders Detection (AVDD) systems using deep learning techniques. These automatic solutions for early diagnosis and monitoring emerge as a promising strategy to address and mitigate the long-term impact of voice disorders. Yet, as it is well known, systems based on deep learning models require a large amount of training data (see, for example, Martinez et al., 2012; Hsu et al., 2018; Wu et al., 2018; Verde et al., 2021; Alhussein & Muhammad, 2018; Harar et al., 2019; Mohammed et al., 2020). This is a major problem in voice disorder detection, as publicly available datasets for this task are scarce. Furthermore, existing databases often suffer from significant class imbalance between pathological speakers and healthy ones. In our prior work (Ribas et al., 2023) we trained an AVDD system using two publicly available datasets: the Saarbruecken Voice Dataset (SVD, see Pützer & Barry, 1997) and the Advanced Voice Function Assessment Database (AVFAD, see Jesus et al., 2017), achieving good overall results despite these limitations.

Several challenges arise when developing AVDD systems. One critical issue is that failing to diagnose a voice disorder (false negative) is much more serious than mistakenly diagnosing a healthy person (false positive). A false alarm typically requires the person to undergo additional tests, which may involve some inconvenience or cost. Failing to diagnose a disorder may result in delayed treatment, potentially resulting in an irreversible condition, a costlier scenario. To address this imbalance, evaluation of system performance should be done using the expected cost (EC) (Ferrer & Ramos, 2025; Maier-Hein et al., 2024) which allows developers to assign different weights to errors based on their relative importance for the task. This metric was used in our prior work (Vidal et al., 2024).

Another key challenge in AVDD, as in any medical task, is to ensure that the system's outputs can be interpreted by medical professionals. Physicians need scores that can be understood as the posterior probability that a patient has a voice disorder given the provided speech sample. Yet, deep learning-based systems often provide very poor posterior probabilities due to the model overfitting the training data, resulting in overconfident posteriors. Vidal et al. (2024) showed that this problem is present in a deep-learning based AVDD system even though the system has outstanding classification performance. Overconfidence of the posteriors can be fixed with a simple calibration step trained with a small amount of data, significantly improving their interpretability and usefulness in clinical practice, as shown by Vidal et al (2024).

In this work, we explore the quality of a voice disorder system across various demographic groups. We show that, despite having outstanding global classification performance and even after calibration, results on some groups are extremely poor. This problem arises from factors such as limited training data for the group or inherent complexities in a group's characteristics. Hence, while a system's overall performance may appear sufficient for practical use, it is crucial to assess whether such performance holds across all demographic groups for which the system will be used.

The bias in performance across groups affects both the quality of the categorical decisions and the quality of the posteriors, severely restricting the interpretability of such outputs. We propose a simple group-wise calibration approach to improve the interpretability of the posteriors, even when the system's decisions are poor. After applying this technique, the posteriors can be trusted to faithfully relay the confidence that the system has in its decision, being appropriately low for groups where the system performance is poor.

Section 2 explains the experimental setup, including datasets, metrics and models. Section 3 analyses metrics and performance of different kinds of calibration. Section 4 studies the distributions of demographic groups defined by metadata and explores ways to improve performance. Finally, Section 5 summarizes the main conclusions of this study.

## 2. Methods

In this section, we outline the experimental setup used to produce AVDD predictions and evaluate their performance. Specifically, we provide a detailed description of the datasets employed, the classification and calibration models implemented, and the performance metrics used to evaluate the systems.

### 2.1 Databases

For this study, the two previously mentioned datasets were employed: AVFAD and SVD. AVFAD, comprising 32 hours of speech recordings in Portuguese, includes 363 healthy and 346 disordered recordings. This data was exclusively used for model training. SVD consists of 1988 phrases in German, with 634 healthy and 1354 disordered recordings. SVD recordings are graded using the GRBAS scales (Hirano, 1981). The GRBAS scale is the most widely used tool for the perceptual assessment of dysphonic voices. It includes five categorical descriptors: Grade (*G*), Roughness (*R*), Breathiness (*B*), Asthenia (*A*), and Strain (*S*), rated on a scale from 0 to 3, where 0 indicates a normal, healthy voice quality, free from any noticeable abnormalities or disorders, implying that the person's voice functions and sounds as expected for the age, gender, and physical condition, and where 3 reflects severe dysphonia. In this work, the Grade ratings for overall dysphonia level were used as an indicator of overall severity of the disorder. SVD was used for training and testing through a cross-validation approach. Table 1 shows the number of speakers per each of the selected groups for the complete SVD dataset.

| Group | Total | Healthy | Disordered | Prior_h | G=0 | G=1 | G=2 | G=3 |
|---|---|---|---|---|---|---|---|---|
| YF | 420 | 318 | 102 | 0.76 | 32 | 30 | 15 | 2 |
| AF | 402 | 44 | 358 | 0.11 | 84 | 83 | 128 | 27 |
| OF | 280 | 19 | 261 | 0.07 | 43 | 25 | 99 | 41 |
| YM | 195 | 137 | 58 | 0.70 | 20 | 6 | 6 | 3 |
| AM | 316 | 99 | 217 | 0.31 | 57 | 35 | 56 | 35 |
| OM | 361 | 16 | 345 | 0.04 | 32 | 40 | 113 | 98 |
| Pooled | 1974 | 633 | 1341 | 0.32 | 268 | 219 | 417 | 206 |

Table 1: Number of healthy and disordered speakers on the SVD dataset for each demographic group and for the pooled data. For disordered speakers, the number for each value of G is also included. *G* is the overall grade of hoarseness, a feature of GRABS. Not all disordered speakers are annotated with GRABS. Hence, the sum of the G=0 through G=3 columns does not necessarily add up to the column Disordered. The fraction of healthy speakers in each demographic group is listed under the *Prior_h* column.

### 2.2 Metrics

In this work, we are interested in evaluating the performance of AVDD systems in ways that ensure that these systems are not only accurate but also fair and reliable for real-world applications. The AVDD task is a binary classification problem consisting of two classes: healthy and disordered. We take disordered as the positive class and healthy as the negative class.

Two of the most commonly-used metrics for AVDD are accuracy (ACC) and Unweighted Average Recall (UAR) (e.g., Mohammed et al., 2020; Huckvale et al., 2021). ACC and UAR, though, do not necessarily reflect the needs of the task since they do not explicitly consider the relative impact of the different types of error. As we explain below, accuracy considers every error as equally severe, while UAR takes each error type to have a cost inversely proportional to the prior probability of the true class of the sample (see Ferrer & Ramos, 2025; Maier-Hein et al., 2024). In the context of voice disorders, though, these assumptions on the cost of errors may not be appropriate. While false positives (false alarms) can potentially lead to unnecessary tests or treatment for healthy individuals they should probably be considered as less costly than false negatives (missed diagnoses), where a disorder can go untreated possibly evolving into a severe pathology. To address this scenario, we use the Expected Cost (EC), a classic metric that generalizes the total error rate by assigning application-dependent costs to each type of error (Maier-Hein et al., 2024; Godau et al., 2023; Ferrer & Ramos, 2025).

To define the *EC* we need a cost matrix with components $c_{ij}$ which represent the cost we believe that the system should incur for making decision $j$ when the true class of the samples was $i$. These costs are highly dependent on the application and should usually be defined in consultation with experts in the specific task of interest.

Given the cost matrix, the empirical estimate of the expected cost over an evaluation dataset can be computed as (Ferrer, 2025)

$$\text{EC} = \sum_{i=1}^{K} \sum_{j=1}^{D} c_{ij} P_j R_{ij} \quad (1)$$

where $R_{ij}$ is the fraction of samples of class $H_i$ for which the system made decision $D_j$ and $P_i$ is the empirical estimate of the prior probability (i.e., the frequency or prevalence) of class $H_i$ in the evaluation data.

For a binary task like AVDD, taking the costs for making a correct decision as zero, the EC reduces to

$$\text{EC} = c_{FN} P_D R_{FN} + c_{FP} P_H R_{FP}, \quad (2)$$

where $R_{FP}$ is the false positive rate, $R_{FN}$ is the false negative rate, $P_D$ is the proportion of samples belonging to a patient with a disordered voice and $P_H$ is the proportion of samples that belongs to a healthy one. For this work, we define the cost of a false positive as one ($c_{FP} = 1$) and the cost of a false negative as three ($c_{FN} = 3$) and refer to this cost as $\text{EC}_3$.

The EC can be normalised to obtain a more interpretable metric by dividing it by the EC for the best naive system, one that has no access to the input sample, producing always the same decision. The normalised EC is given by (Ferrer & Ramos, 2025)

$$\text{NEC} = \frac{\text{EC}}{\min(C_{FN} P_P, C_{FP} P_H)}. \quad (3)$$

If the NEC value is above 1, then the system is worse than a system that has no access to the input samples.

Note that the EC with all costs for errors equal to one reduces to the total error rate (TER) which coincides with $1 - \text{ACC}$. We will call the normalised version of this EC, NTER. Further, if we take the costs to be $C_{FN} = \frac{1}{2P_D}$ and $C_{FP} = \frac{1}{2P_H}$, we obtain the balanced error rate (BER, and NBER its normalised version) which relates to the UAR by $\text{UAR} = 1 - \text{BER}$.

So far, we have described metrics that assess the performance of categorical decisions. Such decisions are made using the posteriors generated by the classifier using a strategy that minimizes the resulting expected cost. The decisions that minimize a specific expected cost are called Bayes decisions. For binary classification, Bayes decisions are made by comparing the posterior for the positive class with a threshold given by $\frac{C_{FP}}{C_{FP}+C_{FN}}$ (Ferrer and Ramos, 2025). This threshold corresponds to 0.5 for accuracy, 0.68 for UAR (for the priors in our dataset), and 0.25 for $EC_3$ (see Table 1 in Vidal et al., 2024).

In the AVDD task, as well as in most medical tasks, it is desirable for the system to also provide a measure of its certainty, given the input sample. Such a measure is given by the posterior probability. Perhaps surprisingly, though, a system may make excellent decisions based on very poor posterior probabilities. This often occurs with deep neural networks since they tend to overfit the training data, resulting in overconfident posteriors (see, for example, Guo et al., 2017). Hence, when planning to provide the posteriors produced by the system to the end user, it is essential to directly assess their quality and not only that of the decisions made with them. To this end, we use the cross-entropy, which is the expected value of the logarithmic loss, a strictly proper scoring rule (PSR, see Gneiting & Raftery, 2012). PSRs offer a rigorous method to evaluate the quality of posteriors. The EC of Bayes decisions is also a PSR, though not a strict one and, hence, it cannot be used to thoroughly assess the quality of the posteriors (Ferrer & Ramos, 2025; Dawid & Musio, 2014). The cross-entropy is defined as follows

$$\text{XE} = -\frac{1}{N} \sum_{t=1}^{N} \log\left(s_{h(t)}(x^{(t)})\right), \quad (4)$$

The where $N$ is the total number of samples in the evaluation set and $s_{h(t)}(x^{(t)})$ is the system's output for sample $x^{(t)}$ for class $h(t)$, the true class of sample $t$. The logarithmic loss which is used to compute the XE heavily penalizes posteriors that are very certain about an incorrect decision. For this reason, it is an excellent default PSR for evaluating the goodness of systems that output posterior probabilities for high-risk applications where we never want to make an incorrect decision with high confidence.

As with the EC, we can normalise the XE by the XE value of a naive system that has no access to the input samples and only knows the prior distribution. In this case, the normalizer is given by $\sum_{i=1}^{K} P_i \log(P_i)$ and, as for the EC, if the normalised XE value is larger than 1 it means that the system is worse than the naive one.

Given a classifier, it is often possible to improve its XE by processing its outputs with a transformation trained for this purpose on some held-out dataset. Such a transform is called a calibrator. The XE obtained after this transform can be taken as a measure of how much class-related information is contained in the system's outputs, i.e., its discrimination performance. While calibration does not improve a system's discrimination, it improves the quality of the posteriors, making them better for interpretation (Ferrer & Ramos, 2025). The difference between the original XE of a system and its XE after calibration is called Calibration Loss and measures how much of the original XE of the system can be reduced by calibration.

For a binary classifier, the best possible monotonic transform – that is, the one that minimizes the XE without changing the ranking of the scores – can be obtained with an algorithm called Pool Adjacent Violators (PAV, see Ayer et al., 1955). PAV, though, is a non-parametric transform that is quite prone to overfitting the data in which it is trained. For this reason, in this work we use a parametric model given by an affine transformation trained to minimize the XE.

### *2.3 Classification and Calibration Models*

The classification model considered for this study is based on the approach used in our previous works (Ribas et al., 2023; Vidal et al., 2024). The primary component is a HuBERT model (Hsu et al., 2021), a transformer-based architecture trained using self-supervised learning to reconstruct masked segments of speech from their surrounding context, thereby learning robust representations of speech features. In this setup, the HuBERT model serves as a feature extractor and is fine-tuned during training to optimize its performance for the AVDD task. The output sequence generated by HuBERT is processed by another transformer-based model, which incorporates a class token to aggregate information across the sequence. A final output layer, with a softmax activation function, generates the posterior probabilities for the two target classes (disordered and healthy speech). The model is trained using cross-entropy loss using a 5-fold cross-validation approach to obtain posteriors on the full dataset. Additional details on the architecture and training process can be found in (Vidal et al., 2024).

For calibration, as mentioned above, we employed one of the most standard calibration approaches, consisting of an affine transformation of the logarithm of the posterior vector. For the binary case, this transform is called Platt Scaling (Platt et al., 1999). The parameters for the transform are trained by minimizing the cross-entropy. See (Ferrer & Ramos, 2025) for detailed information about the procedure. A 10-fold cross-validation approach was implemented to train and evaluate the calibrator taking the posteriors generated by the classifier described above as input.

We begin by examining the effects of global calibration, based on the approach outlined by Vidal et al. (2024), where a single transformation is learned and applied for all demographic groups. Next, we will advocate for a more specialised strategy by training group-dependent calibration transforms. This approach provides a separate calibration transform for each group, potentially addressing group-dependent biases, albeit with a reduction in the amount of data used for training (see Section 3.2). Both the original classification models and those followed by a calibration stage produce posterior probabilities of the classes as their output: $P(H_d|s)$ and $P(H_h|s)$, where $P(H_d|s) + P(H_h|s) = 1$. In this paper, we will show plots of log-odds and of log-likelihood-ratios (LLR), respectively defined as:

$$\text{log-odds} = \log\left(\frac{P(H_d|S)}{P(H_h|S)}\right), \quad (5)$$

and

$$\text{LLR} = \log\left(\frac{P(H_d|S)\, P_h}{P(H_h|S)\, P_d}\right) = \text{log-odds} - \text{log-odds-of-priors}, \quad (6)$$

where $P_h$ and $P_d$ are the prior probabilities of healthy and disordered for the group for which the LLRs are plotted and where $\text{log-odds-of-priors} = \log\left(\frac{P_h}{P_d}\right)$.

## 3. Comparative analysis of metrics and systems calibration

The AVDD system developed in our prior works described above showed excellent results, both in terms of the quality of the categorical decisions and its posteriors. The accuracy stands at approximately 83%, and the normalised cross entropy undergoes a reduction from 1.80 to 0.65 after global calibration. But, as we will see, despite these overall positive results, a detailed analysis of different groups defined by gender and age reveals large differences in performance across groups, with performance in many groups being close to or worse than random.

To conduct a comprehensive study of the system performance, we divide the speakers based on both their gender and age, based on the work by Hazan (2016). This information is provided in the metadata for both datasets. We aim for the groups to be as homogeneous as possible while having enough samples to estimate performance and train calibration models. Consequently, we use six groups, each comprising three age ranges per gender annotated in the datasets (male and female). The age classification is the following: young (14 to 30 years old), adults (31 to 55 years old), and older adults (56 years old and above). We call the groups Y for young, A for adults, and O for older age, followed by M or F for male and female. Samples from people younger than 14 years old were discarded from our experiments since these samples are scarce and often not representative of the adult population.

In the section below, we discuss the problems with the standard metrics used in the literature for this task, accuracy or UAR computed over the full dataset, and argue in favour of the NEC and NXE computed as averages over demographic groups. We then show a comparison of performance for three systems: the original system without calibration, a system with global calibration, and, finally, a system with group-wise calibration showing that even after group-wise calibration, the performance on some demographic groups is quite poor. Finally, we analyse various factors that might explain this phenomenon.

### 3.1 Comparison of metrics

The bar plots in Fig. 1 show ACC, NTER, UAR, and NBER for each of the demographic groups and for the full test set (we call it pooled, the commonly used value). The metrics in this plot correspond to the uncalibrated posteriors, with decisions made by thresholding the posterior for disordered at 0.5 for ACC and NTER and at 0.68 for UAR and NBER. We also include the average value of the metrics across groups.

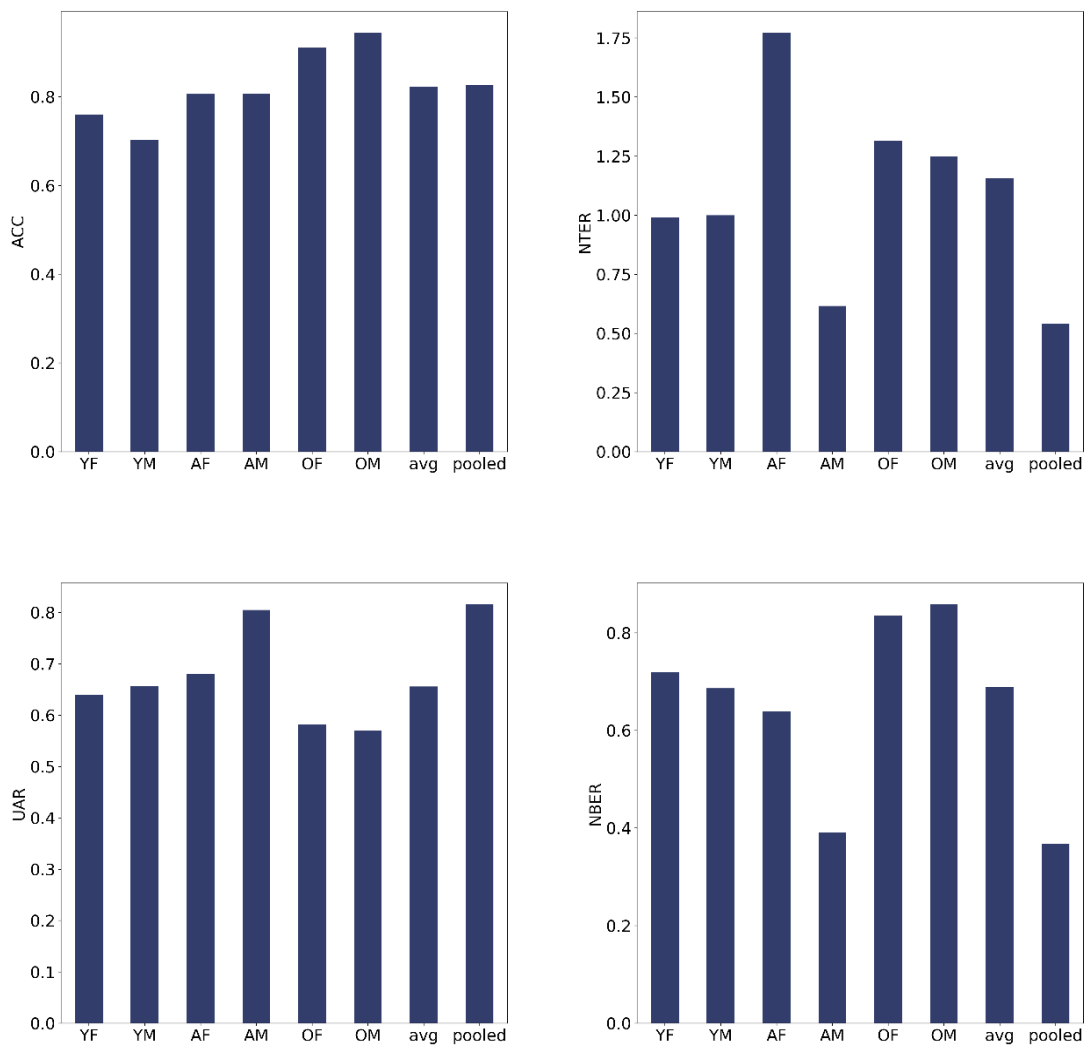

Figure 1: Top left: Accuracy values for all six groups, the average between them and the accuracy on the pooled set. Top right: Equivalent plot for NTER, which is related to ACC as explained in Section 2.2. Bottom left: Equivalent plot for UAR. Bottom

right: Equivalent plot for NBER, related to UAR as explained in Section 2.2. Values correspond to the uncalibrated posteriors. The avg bar is the average of the bars across groups.

Subplots on the left of Fig.1 show that the pooled ACC and UAR values are reasonably high (0.83 for ACC and 0.82 for UAR) but the performance is uneven across groups. In the case of ACC, the worst performance occurs on the younger groups (reaching a minimum of about 0.7 for males) while for UAR the worst performance is found on older people (0.58 for both genders). Since ACC weights each sample equally and the size of the groups is similar, the average and pooled ACC have similar value. On the other hand, the average UAR is significantly worse than the UAR of the pooled set.

Subplots on the right of Fig.1 show NTER and NBER. Note that, contrary to ACC and UAR, for these metrics, lower values are better. While ACC is directly related to NTER and UAR directly related to NBER, as explained above, we can see a notable difference in behaviour between the metrics in each pair due to the normalization process, which is done using the priors specific to each group. This normalization makes the metrics across groups comparable to each other since it results in a value that is relative to the best possible naive classifier for each group. For example, for the AF group, while ACC appears satisfactory at first glance, the NTER value is quite high indicating that the system is not informative: the prior probability for this group is actually higher than the achieved accuracy, meaning that a naive system that always predicts the majority class would perform better than our classifier. This example clearly illustrates why using normalised metrics is essential: they provide a more interpretable assessment of the system's performance by accounting for the underlying class priors, whereas unnormalised metrics like ACC and UAR can be misleading when the dataset is imbalanced. Note also that NTER and NBER present a larger difference between pooled and average values than ACC and UAR, with the high average value correctly reflecting the fact that the system works poorly on some groups. These results illustrate the importance of selecting the evaluation metric to adequately reflect the performance that the system will have in practice.

Fig. 1 shows a somewhat surprising trend. For three of the metrics, UAR, NBER and NTER, the pooled value is comparable to that of the group with the best performance. All other groups show a poorer performance that does not seem to affect the pooled value. We hypothesized that his phenomenon could be explained by the fact that age and gender on their own are very good predictors of the class since the class priors depend heavily depend on those factors (see Table 1). To test this statement, we trained a Random Forest model that uses only gender and age range to predict the class, using a cross-validation approach with 10 folds. The Random Forest model reaches an accuracy and a UAR of 83%, in both cases a performance only slightly worse than that of the AVDD system (see Table 2 from the Appendix for detailed results). This indicates that the AVDD system provides minimal additional predictive power beyond what can be achieved using demographic data alone, a fact that is not apparent in the pooled ACC and UAR, the metrics commonly used in the literature. In contrast, this conclusion is properly reflected in the value of the average NTER and NBER which reach values close to 1.0, the performance of a naive system.

The average NTER and NBER overcome some of the problems with the standard metrics, the pooled ACC and UAR. Yet, these metrics still have the disadvantage of relying on costs that may not appropriately reflect the requirements of this task. For this reason, in the next sections we show results in terms of NEC$_3$, a metric specifically designed to capture the relevant costs for this application. Further, we also show NXE, which, as explained above, reflects the overall performance of the posteriors produced by the systems.

## 3.2 Comparison of systems

In this section we compare three systems: the one we call the 'raw', which has no calibration stage, the one calibrated with a global transformation, and the one where each group is calibrated separately, which we call 'group-wise'.

Fig. 2 shows the NEC$_3$ (left) and NXE (right) by group, averaged and for the pooled set for each of the systems. For both metrics, we also include the value after calibration with an affine model using the test data itself for training the calibrator parameters. This value is shown as black horizontal line within each bar and indicates the best value that can be obtained by calibration on those samples. If the black line is close to the top of the bar, it means that no further improvement can be achieved by calibrating with an affine model.

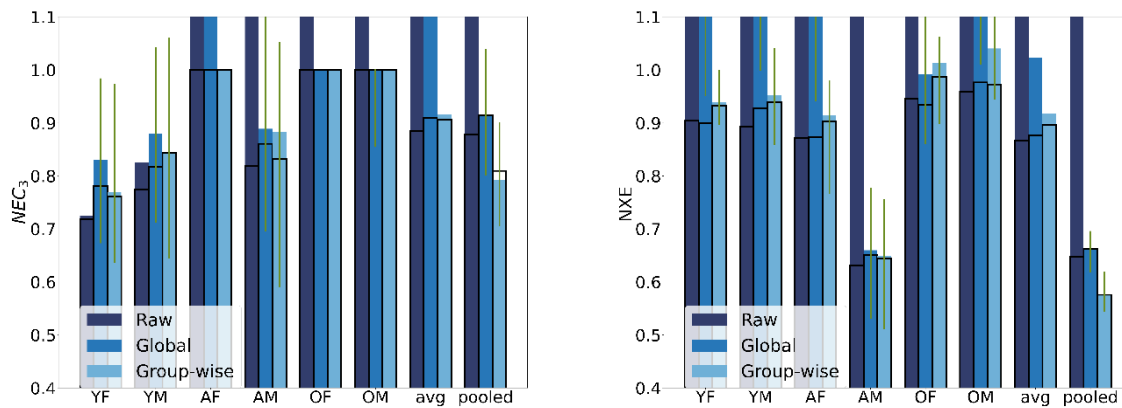

Figure 2: NEC$_3$ (left) and NXE (right) values for all six groups, and the pooled dataset and the average across groups, for the raw, the globally calibrated, and the group-wise calibrated systems. Horizontal lines within the bars indicate the metric value after calibrating on the evaluation data for each group. Uncertainty bars are calculated by bootstraping (Keller et. al., 2005) train and test data.

As for the metrics discussed in the previous section, NXE and $NCE_3$ show a large disparity across groups. For the raw and globally-calibrated systems, several groups have a performance that is worse than that of a naive system. Also, as for the metrics discussed above, the pooled value of these metrics is quite optimistic compared to the per-group results. For example, the NXE of the globally calibrated system is 0.65, suggesting that the system produces reasonably good posteriors. Yet, this pooled result hides the fact that only for the AM group the system performance is good. For all other groups, the cross-entropy value is significantly higher, with values well above 1.0 for several groups. As a consequence, the average NXE of the globally calibrated system is around 1.0 correctly diagnosing that, on average, the system provides very little information beyond that of a random system.

Group-wise calibration gives an improvement over global calibration – in some cases, a very large one. Further, we see that, for most groups, the NXE after group-wise calibration is very close to the one that can be obtained after calibration with the test data (the black horizontal line is close to the top of the bar). This implies that the calibration model learned on the training data generalizes well to the test data. Importantly, group-wise calibration not only improves the quality of the posteriors, as indicated by NXE, but also the quality of the categorical decisions, as indicated by a lower average $NEC_3$. Nevertheless, even after group-wise calibration we can see that all groups except one (AM) have an NXE above 0.9. Calibration cannot solve the poor discrimination of the system in those groups. Notably, though, despite having a high NXE, the NEC$_3$ is reasonably good in the younger groups, meaning that the decisions on those groups are significantly better than those of a naive system.

An important implication arising from these detailed per-group analyses is that they precisely define the scope and limitations of automated systems intended for diagnosis or assessment of disorders. Specifically, our findings suggest that a diagnostic test based on analysing a single phrase from a subject may indeed be reliable within a certain age range but might lack sufficient accuracy for individuals outside of that range, who may require more comprehensive and nuanced diagnostic methods.

In the next section, we provide further analyses of the systems' outputs in an attempt to understand these results.

## 4. Analysis and discussion

### 4.1 Score distributions

Fig. 3 shows the density distribution of LLR for each demographic group for healthy (right) and disordered (left) cases for the globally and group-wise calibrated systems. The curves are normalised so that the area under each distribution equals 1. We show the distribution of LLRs instead of log-odds since the LLRs are independent of the class priors, facilitating the comparison across groups. An LLR larger than 0 indicates that, according to the system, the observed sample has a higher likelihood of coming from an individual with a voice disorder than from a healthy individual.

The top right plot shows that, for the globally-calibrated system, for younger age groups the LLRs for healthy individuals are mostly negative indicating that the system correctly assigns them a larger likelihood of being healthy than of having a voice disorder. However, for adult middle-aged individuals, especially females, the distribution of LLR for healthy individuals becomes bimodal, with one of the peaks on the positive side indicating that the system incorrectly assigns a larger likelihood of disorder than of healthy for those cases. For older-age groups, the distribution is completely skewed to the right, indicating that the system assigns a larger likelihood of disorder than of healthy for most healthy older individuals.

The top left plot in Fig. 3 shows the density distributions for instances labelled as disordered for the globally-calibrated system. In this case, we can see that most younger individuals and a large number of middle-aged adults with a voice disorder are incorrectly assigned low LLR values.

The middle row in Fig. 3 shows the corresponding plots for the group-wise calibrated system. We can see that the range of the scores is reduced and closer to zero for that system compared to the globally-calibrated system (note the change of scale in the x-axis). This reduction in scale shows that the globally calibrated system is overconfident, producing LLRs that are too large, not accurately reflecting the uncertainty of the system within each group. The group-wise calibrated LLRs, on the other hand, are close to 0 for most groups, correctly indicating that the system is quite uncertain about the class of the sample. Importantly, note that calibration does not improve the system discrimination. In particular, it cannot do anything about the double modes observed in some of the groups.

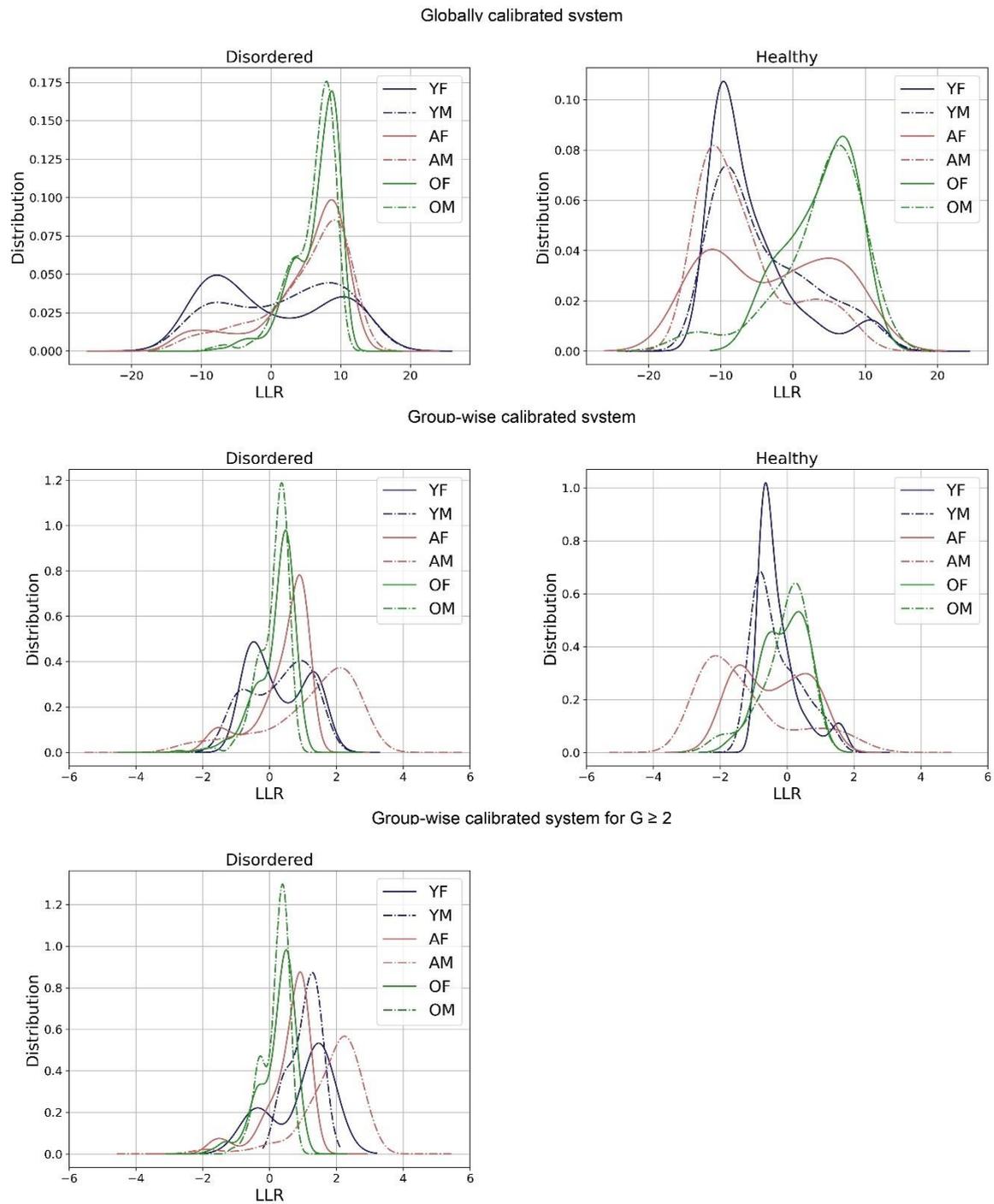

Figure 3: Distribution of LLRs for the globally calibrated system (above) and the group-wise calibrated system (middle) for disordered (left) and for healthy (right) individuals, for each demographic group. We also include a plot (bottom) with only the disordered individuals with G larger than or equal to 2.

Finally, the bottom plot in Fig. 3 shows the LLRs for the disordered individuals with a G value larger or equal to 2. We can see that the fraction of negative LLRs is smaller for this subset of the data, indicating that those difficult samples correspond to individuals with G smaller than 2, i.e., with a low-severity pathology. We will comment more on this phenomenon in section 4.3.

### *4.2 Class- and group-balanced training*

Given the marked difference in class priors and number of speakers for the different age and gender groups (see Table 1), we hypothesised that the performance in some of the groups may be improved by training the system with balanced data. It is particularly suggestive that Fig. 3 shows problematic distributions (bimodal or centred on the wrong side) and thus larger rates of misclassified instances for groups with a smaller number of speakers of the corresponding class (YF and YM for disordered and AF, OF and OM for healthy). In an effort to solve this problem, we trained a new system where batches are created with an equal number of speakers of each demographic group and each class by randomly sampling from the corresponding data. The model trained in this way is not exposed to the true class priors in each group, seeing a 0.5 chance of each class for each group within each training batch.

Results for the balanced-training system are shown in Table 2 in the Appendix. As a summary, Fig. 4 shows the false positive rate ($R_{FP}$) and false negative rate ($R_{FN}$) for each group, for the group-wise calibrated system and the balanced group-wise calibrated system. $R_{FP}$ represents the proportion of negative instances (those from healthy individuals) classified as disordered, and $R_{FN}$ the proportion of positive instances (those from individuals with a voice disorder) classified as healthy, using the Bayes threshold for NEC3. As could be anticipated by looking at Fig. 3, for younger individuals, the false negative rate is relatively high, indicating that many disordered individuals are incorrectly classified as healthy, while the false positive rate is low. For adult individuals, particularly men, the two rates are more balanced. Finally, older individuals have a higher false positive rate and a lower false negative rate, meaning that many healthy individuals are mistakenly classified as having a disorder. The balanced system reflects a very similar behaviour, leading to no significant improvement in performance.

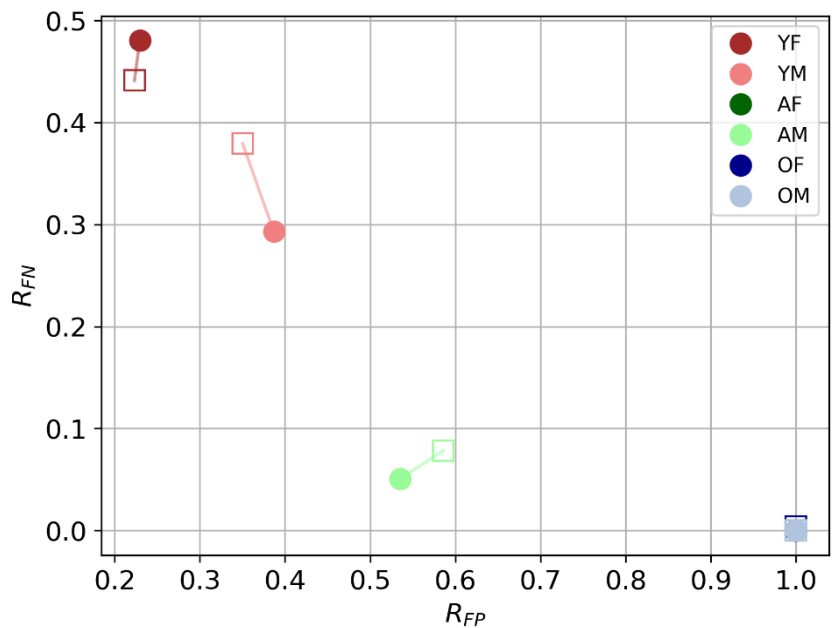

Figure 4: Percentage of misclassified samples of each class, with false positive rate ($R_{FP}$, healthy individuals labelled as disordered) on the x-axis and the false negative rate ($R_{FN}$, individuals with a disorder labelled as healthy) on the y-axis, for the original system trained with batches sampled randomly from the training data (filled circles) and the system trained with batches balanced by age, gender and class (empty squares), using the Bayes threshold for NEC3. In both cases, the results

correspond to the systems after group-wise calibration. Markers for the AF and OF groups are behind those for the OM group.

These results suggest that the difference in class priors across groups is not a significant cause of the poor performance in the younger and older groups. If the system was learning to predict that younger people should be labelled as healthy and older people as disordered based on the priors observed in the training data, this effect would disappear when training with balanced data where the class priors are 0.5 for every group. Yet, the balanced system has the same trend as the original one. Since they cannot discriminate one class from the other, both systems after group-wise calibration simply tend to classify every sample from a group as the majority class in that group. This is the best they can do with uninformative scores. As we discuss in the next section, other factors which would not be addressed by balancing the batches may explain the poor performance of the original system on most groups.

### *4.3 Severity of the cases* and vox senilis

The lack of improvement on the average metrics obtained from the experiment described above where the model is trained with balanced data in an effort to break the correlation between age and class, suggests that other factors may be contributing to the poor performance on most of the demographic groups. Here we explore two important factors that might impact the system performance: the severity of the disorder and the aging of the voice in older individuals.

The left plot in Fig. 5 shows the density distributions for the log-odds for the different $G$ levels where a strong – perhaps obvious – trend is observed: cases with higher G levels are easily identified as disordered by the system with most samples being on the right of the NEC3 threshold (which corresponds to $\log(1/3) \sim -1.1$ for log-odds), while cases with lower G levels are more frequently on the left of the threshold and, hence, labelled as healthy.

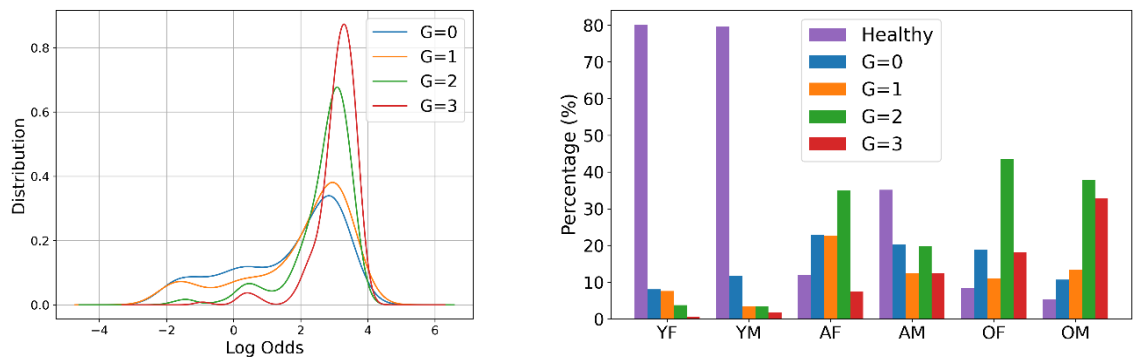

Figure 5: Left: Plot showing the density distributions of log odds for the group-wise calibrated system, for disordered instances with annotated G, for the different levels. Right: Plot bar showing the percentage of samples for each group for healthy and disordered speakers for each level of the grade G.

To understand how this phenomenon affects each of the groups, we analyse the number of samples per G level per group, shown in the right plot of Fig. 5. The corresponding numerical data can be found in Table 1. We can see that while older individuals with a disorder tend to have higher G levels, younger individuals exhibit lower G levels. This explains the high false negative rate for the younger groups. For the adult age-range, we can see that females have a smaller percentage of G=3 compared to males, which may explain the poorer performance in that group compared to adult males.

Fig. 3 (bottom right) in Section 4.1 shows the distribution of LLRs for disordered individuals with G larger or equal to 2 where we can see that the number of samples with a negative LLR is reduced, especially for younger individuals. As a summary of the impact of low-G samples on the results, Fig. 6 shows the false positive rate ($R_{FP}$) and false negative rate ($R_{FN}$) for each group, for the group-wise calibrated system for all samples and only for those with G larger o equal to 2. As before, we use the Bayes threshold for NEC3. For younger males, the $R_{FN}$ reduces almost to zero after discarding the low severity cases while for younger females it reduces by 36%.

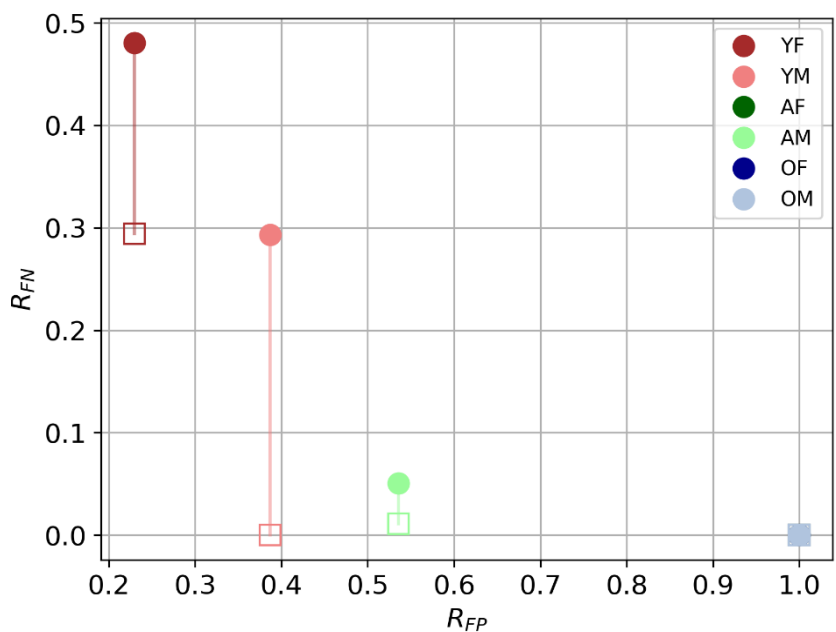

Figure 6: Same plot as Fig. 4, for the all samples in each group and after discarding instances with G lower than 2, for the group-wise calibrated system. Decisions are made using the Bayes threshold for NEC$_3$.

Turning to the older population, we hypothesized that the poor performance on these groups may be due to the fact that their voices often sound like the voices of young or middle-aged adults with a disorder. Hence, a model trained with all ages may tend to confuse older healthy voices with younger disordered voices. A model trained only with older voices, though, would not have this problem. To test this hypothesis, we trained a model using only the older groups. Notably, we found no significant change in performance on the older individuals between this model and the original one trained on all data. This suggests that this hypothesis is incorrect or that the amount of data available to train a model specific to the older individuals is not sufficient to obtain a robust system. In future works, we will further explore this issue.

Another hypothesis to explain the poor performance on older groups is that these voices are likely not well represented by the HuBERT model that we are using as feature extractor since this model was trained using Librispeech (Panayotov et al., 2015), a dataset consisting of audio books where older voices are extremely rare. This issue may be impacting the performance of our model on this age range. In the future, we plan to explore the use of models fine-tuned with speech from older individuals.

## 5. Conclusions

In this work, we conducted a comprehensive analysis of an Automatic Voice Disorders Detection (AVDD) system, revealing critical insights about both system performance and evaluation methodologies. In the AVDD literature, systems are usually evaluated based on metrics like accuracy or unweighted average recall, computed over the pooled dataset. In this work we showed that these metrics are inadequate for the task for several reasons. First, since they are computed over the pooled dataset, they are strongly affected by correlations between the prevalence of voice disorders and demographic characteristics of the speakers, particularly because the prevalence is very different across groups. Due to these differences, a system that bases its predictions on the priors for each demographic group would appear to be working well based on the pooled metrics. Notably, our analysis showed that a system that detects a disorder based only on age and gender of the person, without access to their speech, gives a similar pooled accuracy than our system of around 83%, significantly better than the performance of a system that always selects the majority class on the pooled data, which is 70%. The pooled metric, then, does not adequately reflects the fact that the performance of the system is only marginally better than that of a naive system that only knows the prevalence of voice disorders for each demographic group.

To overcome this problem of the pooled metrics, we propose to use normalized metrics averaged over demographic groups. The normalization ensures that the baseline value for all groups is the same so that the average is more meaningful. We showed that, unlike pooled metrics, average of normalized metrics adequately reflects the fact that the system is only slightly better than a random system. Another problem with classic AVDD evaluation procedures is that the metrics that are commonly used, like accuracy or average recall, do not adequately reflect the needs of the task. We propose to use of a normalized expected cost (NEC) metric averaged across demographic groups, with costs determined such that failing to detect a voice disorder is penalized more heavily than incorrectly detecting a disorder in a healthy person. In our work, we use a 3 to 1 cost ratio. In practice, the costs should be determined specifically for the application of interest in consultation with medical practitioners.

Finally, another problem with the classic metrics is that they do not assess the goodness of the posteriors produced by the system, only that of the categorical decisions made with them. Having good posteriors is essential in tasks where the end user needs to understand the confidence with which the system made its decision. To assess the goodness of posteriors we propose the use of the normalized cross-entropy, NXE, again averaged over demographic groups. We argue that the NXE, in addition to the NEC, reported by demographic group and on average, provides a faithful and comprehensive assessment of the quality of an AVDD system.

In addition to proposing new metrics for evaluation of AVDD systems, we analyse our DNN-based system in depth, in an effort to explain its poor performance on some demographic groups -- a phenomenon that was uncovered by the proposed metrics. We first attempted to train a system with data balanced by class for each age and gender group. Such a system is not affected by the correlation between prevalence of voice disorders and demographic characteristics, since the correlation is broken by the balancing approach. Yet, the performance of this system is still quite poor on the younger and older individuals, suggesting that other factors are at play. In this work we identified two such factors. For younger speakers with voice disorders, the prevalence of low GRBAS G scores partially explains their frequent misclassification as healthy since it is understandably harder to detect a voice disorder for low severity cases. On the other hand, the systematic misclassification of healthy older speakers as disordered may be attributed to the natural aging of voice (vox senilis) combined with the fact that the HuBERT model used by our system as feature extractor was trained primarily on younger voices from the LibriSpeech dataset.

Finally, we showed that, in terms of quality of posteriors, the original uncalibrated system is quite poor. This is a general phenomenon in DNN-based systems, since they are likely to overfit due to their large number of parameters, resulting in overconfident posteriors. To solve this problem, we proposed a group-wise calibration approach and showed that the quality of the posteriors is greatly improved after this transformation.

These findings have important implications for the development and deployment of AVDD systems in clinical settings. First, they emphasize the critical importance of evaluating system performance using an approach that addresses the needs of the task and provides a comprehensive assessment of performance, both in terms of quality of decisions and of posteriors, across demographic groups. Second, they highlight the need for careful consideration of age-related voice characteristics in both feature extraction and model development. Future work should focus on developing more robust feature extraction methods that better account for age-related voice variations, particularly by incorporating more diverse training data that includes older voices.

We recommend that developers of medical AI systems implement group-wise analysis as a standard practice when demographic metadata is available. This approach not only provides a more nuanced understanding of system performance but also helps ensure that the technology can serve all patient populations effectively and equitably.

## Appendix

For completeness, Table 2 shows the values of the metrics for the different groups, the average and the pooled sets for the different approaches discussed. Values in this table match those of Fig. 1.

| Metric | group | YF | AF | OF | YM | AM | OM | avg | pooled |
|---|---|---|---|---|---|---|---|---|---|
| | count | 420 | 402 | 280 | 195 | 316 | 361 | - | 1974 |
| ACC | Raw | 0.76 | 0.81 | 0.91 | 0.70 | 0.81 | 0.94 | 0.82 | 0.83 |
| | Global Cal | 0.74 | 0.83 | 0.93 | 0.65 | 0.83 | 0.94 | 0.82 | 0.83 |
| | Group-wise Cal | 0.78 | 0.89 | 0.93 | 0.73 | 0.84 | 0.96 | 0.85 | 0.86 |
| | Balanced Group-wise Cal | 0.77 | 0.89 | 0.93 | 0.73 | 0.84 | 0.96 | 0.85 | 0.86 |
| | Metadata-based system | 0.76 | 0.89 | 0.93 | 0.70 | 0.69 | 0.96 | 0.82 | 0.83 |
| UAR | Raw | 0.64 | 0.68 | 0.58 | 0.66 | 0.80 | 0.57 | 0.66 | 0.82 |
| | Global Cal | 0.62 | 0.66 | 0.61 | 0.62 | 0.80 | 0.50 | 0.63 | 0.82 |
| | Group-wise Cal | 0.64 | 0.71 | 0.59 | 0.66 | 0.80 | 0.57 | 0.66 | 0.85 |
| | Balanced Group-wise Cal | 0.68 | 0.73 | 0.57 | 0.66 | 0.80 | 0.53 | 0.66 | 0.85 |
| | Metadata-based system | 0.52 | 0.50 | 0.50 | 0.53 | 0.50 | 0.50 | 0.50 | 0.80 |
| NTER | Raw | 0.99 | 1.77 | 1.32 | 1.00 | 0.62 | 1.25 | 1.16 | 0.54 |
| | Global Cal | 1.07 | 1.52 | 1.05 | 1.19 | 0.54 | 1.25 | 1.10 | 0.53 |
| | Group-wise Cal | 0.92 | 1.00 | 1.05 | 0.91 | 0.52 | 1.00 | 0.90 | 0.44 |
| | Balanced Group-wise Cal | 0.96 | 1.00 | 1.10 | 0.91 | 0.52 | 1.00 | 0.92 | 0.50 |
| | Metadata-based system | 0.99 | 1.00 | 1.00 | 0.98 | 1.00 | 1.00 | 1.00 | 0.53 |
| $NEC_3$ | Raw | 0.73 | 4.16 | 1.58 | 0.82 | 1.27 | 1.88 | 1.74 | 1.11 |
| | Global Cal | 0.77 | 1.93 | 1.00 | 0.83 | 0.86 | 0.94 | 1.06 | 0.88 |
| | Group-wise Cal | 0.72 | 1.00 | 1.00 | 0.76 | 0.87 | 1.00 | 0.89 | 0.77 |
| | Balanced Group-wise Cal | 0.71 | 1.00 | 1.15 | 0.83 | 1.10 | 1.00 | 0.96 | 0.81 |
| | Metadata-based system | 0.96 | 1.00 | 1.00 | 1.19 | 1.00 | 1.00 | 1.02 | 1.00 |
| NBER | Raw | 0.72 | 0.64 | 0.84 | 0.69 | 0.39 | 0.86 | 0.69 | 0.37 |
| | Global Cal | 0.75 | 0.68 | 0.79 | 0.77 | 0.40 | 1.00 | 0.73 | 0.37 |
| | Group-wise Cal | 0.71 | 0.59 | 0.81 | 0.68 | 0.40 | 0.86 | 0.67 | 0.30 |
| | Balanced Group-wise Cal | 0.65 | 0.53 | 0.87 | 0.67 | 0.39 | 0.95 | 0.68 | 0.30 |
| | Metadata – based system | 0.96 | 1.00 | 1.00 | 0.94 | 1.00 | 1.00 | 0.98 | 0.41 |
| NXE | Raw | 3.17 | 3.71 | 2.11 | 2.99 | 1.70 | 2.19 | 2.64 | 1.80 |
| | Global Cal | 1.08 | 1.09 | 0.99 | 1.10 | 0.64 | 1.13 | 1.01 | 0.65 |
| | Group-wise Cal | 0.92 | 0.90 | 0.99 | 0.91 | 0.64 | 0.99 | 0.89 | 0.57 |
| | Balanced Group-wise Cal | 0.90 | 0.79 | 0.99 | 0.91 | 0.67 | 1.01 | 0.88 | 0.56 |
| | Metadata-based system | 1.01 | 1.01 | 1.01 | 1.00 | 1.00 | 1.02 | 1.00 | 1.80 |

Table 2: Values for ACC, UAR, NTER, $NEC_3$, NBER and NXE, for the raw, the globally calibrated and the group-wise calibrated systems, for the balanced group-wise calibrated system and for a metadata-based system (a random forest classifier); for each of the six groups, averaged and pooled.

## Data availability statement (DAS)

The two datasets used in this work are publicly available upon request to their authors.